\documentclass[preprint,showpacs,preprintnumbers,amsmath,amssymb]{revtex4}
\usepackage{graphicx}
\begin{document}

\title{Entanglement properties of a two spin-one particle system under a Lorentz transformation}

\author{Esteban Castro-Ruiz, and Eduardo Nahmad-Achar}

\affiliation{Instituto de Ciencias Nucleares,
Universidad Nacional Aut\'onoma de M\'exico \\
Apdo. Postal 70-543, Mexico D. F., C.P. 04510}

\date{\today}


\begin{abstract}
\noindent We analize the entanglement change, under a Lorentz transformation, of a system consisting of two spin-one particles, considering different partitions of the Hilbert space, which has spin and momentum degrees of freedom. We show that there exists a complete set of states of the spin subspace in which the entanglement change of any state in the set is zero for all partitions and all values of the Wigner angle. Moreover, these states only change by a global phase factor under the Lorentz boost. Within this basis, maximally entangled invariant states, interesting for quantum information purposes, are explicitly obtained. On the other hand, the entanglement in the particle-particle partition {\it is} Lorentz invariant, thus protecting the consistency of quantum correlations and teleportation results. We show how our results may be generalized to arbitrary spin.
\end{abstract}



\maketitle

\section{Introduction}

\noindent The instantaneous collapse of a two-particle quantum state vector has no covariant meaning, since the time-ordering of two different measurements depends on the reference frame for spatially separated events. In order to analyze this problem in a concrete situation, it is convenient to study entanglement from a relativistic point of view \cite{Peres}. This analysis is also important for the study of relativistic quantum infomation tasks \cite{Gingrich, Terno}. In this work, we analyze the entanglement change, after a Lorentz boost, of a system composed by two spin-one massive particles in an EPR-like situation \cite{EPR}. We compare our results with similar analyses done by Friis et al., and Jordan et al., for spin-$1/2$ particles \cite{Friis,Jordan}. As well as these authors, we consider the particle's momentum as a discrete two-level variable, and calculate the entanglement change, with respect to different decompositions of the Hilbert space, caused by a Lorentz boost in a fixed direction. We address the question of the dependence of the entanglement change on the initial spin entanglement and argue that, instead of the initial entanglement, quantum superposition plays the key role in the initial entanglement-final entanglement relation. We focus on initial spin states for which no entanglement change occurs at all, and show that, for the situation under study, such states form a complete basis of the two-particle spin Hilbert space. Within this basis, maximally entangled invariant states, interesting for quantum information purposes, are explicitly obtained. On the other hand, since there are elements of this basis that suffer a distinct global phase change under the boost, an arbitrary linear combination of them will {\it not} in general be Lorentz-invariant. We generalize our results to higher spins. 

In Section \ref{two} we introduce, for the sake of competeness, the relativistic transformation of momentum and spin states, while Section \ref{three} gives a description of the problem under study. The results are presented and discussed in Section \ref{four}, where we also show the existence of a complete set of invariant states the two-particle spin space. Conclusions are presented in Section \ref{five}.   




\section{Lorentz transformations and Wigner rotations}
\label{two}
\noindent In this section we present the relativistic transformation of a single-particle state with momentum and spin degrees of freedom. We use units such that $c=\hbar=1$. 

The four-momentum of a particle in its own rest frame is given by 
\begin{equation}
k = \begin{pmatrix}
	m \\
	\mathbf{0}
 \end{pmatrix},
\end{equation}
\noindent where $m$ is the rest mass of the particle. We label the four-vector components with the set of indices $\{0,1,2,3\}$, and use boldface letters to denote three-vectors. The quantum state of the particle in the rest frame is denoted by $\vert k \, \sigma\rangle$. This state has a well-defined linear four-momentum, $k$, and a well-defined spin projection along the $z$-axis, $\sigma$. In our case, $\sigma \in \{1,0,-1\}$. Let $L_p$ denote the standard Lorentz boost that takes the rest frame four-momentum $k$ to the four-momentum $p$. The four-momentum eigenstates are defined in terms of the rest-frame state and the standard boosts as follows \cite{Weinberg}:
\begin{equation}
\label{deffinition}
\vert p, \, \sigma \rangle = U(L_p) \vert k, \, \sigma \rangle,	
\end{equation}
\noindent where $U$ is a unitary representation of the Poincar\'e group on the complete momentum-spin Hilbert space. These states form a complete set and satisfy
\begin{align}
P^\mu\, \vert p, \, \sigma \rangle =  p^\mu\, \vert p, \, \sigma \rangle, 
\end{align}
where $P^\mu$ is the $\mu$-th component of the four-momentum operator, and $p^\mu$ stands for the $\mu$-th component of $p$. \\

Now consider an arbitrary Lorentz transformation $\Lambda$. The action of $\Lambda$ on  the four-momentum eigenstate $ \vert p \, \sigma\rangle$ is given by
\begin{align*}
U(\Lambda)\vert p,\,\sigma\rangle =& \, U(\Lambda)U(L_p) \vert k, \, \sigma \rangle\\
=&\, U(L_{\Lambda_p})U(L^{-1}_{\Lambda_p}\Lambda L_p)\vert k, \, \sigma \rangle\\
=&\, U(L_{\Lambda_p})U(W(\Lambda,p))\vert k, \, \sigma \rangle.\\
\end{align*}
Here we have defined the Wigner rotation as
\begin{equation}
\label{Wigner rotation}
W(\Lambda,p) = L^{-1}_{\Lambda_p}\Lambda L_p 
\end{equation}
It is clear that $W(\Lambda,p)$ is a pure rotation, since it leaves the four-momentum $k$ invariant: $L^{-1}_{\Lambda_p}\Lambda L_p k = k$. The rotation angle $\Omega$, called the Wigner angle, is a function of the magnitudes of the rapidities $\mathbf{\eta}$ and $\mathbf{\omega}$ that define, respectively, the boosts $L_p $ and $\Lambda$, and under a suitable choice of setting (see below) may be written as \cite{Alsing}:
\begin{equation}
\label{Wigner angle}
\tan\Omega = \frac{\sinh \vert \mathbf{\eta}\vert \sinh\vert\mathbf{\omega}\vert}{\cosh\vert \mathbf{\eta}\vert+\cosh\vert \mathbf{\omega}\vert}.	
\end{equation}
Thus, under an arbitrary Lorentz transformation $\Lambda$, the momentum is changed from $p$ to $\Lambda p$ and the spin part of the state transforms under the action of the rotation group, since, by the definition of the four-momentum eigenstates (Eq.(\ref{deffinition})), the spin label remains unchanged after the standard boost $L_{\Lambda_p}$. Hence, we have
\begin{equation}
\label{transformation}
U(\Lambda)\,\vert p\,\sigma\rangle=\sum_{\sigma^{\prime}}\,D^{(j)}_{\sigma^\prime\,\sigma}(W(\Lambda,\,p))\vert\Lambda p\,\sigma^{\prime}\rangle,	
\end{equation}
where $D^{(j)}$ is a spin-$j$ representation of the rotation group. In our case, $j$ = 1.

\section{Description of the problem}
\label{three}
\noindent We consider an EPR-like situation 
with two spin-1 particles propagating in opposite directions with respect to each other, and compare the entanglement of the system before and after a Lorentz boost on the state. We take the direction of propagation to be the $z$-axis. One of the particles is supposed to be under control of one observer, Alice, while the other is supposed to be under control of another spatially separated observer, Bob. The entanglement is calculated with respect to different partitions of the complete Hilbert space, which is a tensor product of Alice's and Bob's Hilbert spaces. Each of the single-particle spaces is composed by spin and momentum degrees of freedom, so that the complete Hilbert space can be written as a tensor product of four physically distinct subspaces:
\begin{equation}
H = H^{(A)}_p \otimes H^{(A)}_s \otimes H^{(B)}_p \otimes H^{(B)}_s.
\end{equation}
Here, the superscrips $(A)$ and $(B)$ denote Alice and Bob subsystems, while the subscripts $p$ and $s$ denote the momentum and spin degrees of freedom, respectively. Following \cite{Friis}, we calculate the entanglement change of the state, after a Lorentz boost, with respect to three different partitions of $H$: {\it i)} the $A$ vs. $B$ partition, that is, the partition formed by Alice's and Bob's subsystems; {\it ii)} the $p$ vs. $s$ partition, obtained by tracing over the momentum or spin degrees of freedom of the whole state in $H$; and {\it iii)} the entanglement change with respect to the partition formed by one given subspace opposed to the other three; we shall call this the $1$ vs. $3$ partition. 

We choose the initial state $\vert \psi \rangle$ to be a pure and separable state in $p$ and $s$, i.e.
\begin{equation}
\label{initial state}
\vert\psi\rangle=\vert p\rangle\otimes\vert s\rangle,
\end{equation}
where $\vert p\rangle \in H^{(A)}_p \otimes H^{(B)}_p$ and $\vert s \rangle \in H^{(A)}_s \otimes H^{(B)}_s$. Furthemore, we consider only sharp-momentum distributions, i.e. states where the momentum of each particle is concentrated around a definite value that can be $p_+$, for propagation along the positive direction of the z-axis, and $p_-$, for propagation along the opposite direction. For a study of wave packets in the spin $1/2$ case, see \cite{Gingrich}, while a wave packet analysis for the photonic case may be found in \cite{Peres photons, Bradler}. The states in this work are considered to be normalized in the sense that $\langle p_i \vert p_j \rangle = \delta_{ij}$, where the subscripts $i$ and $j$ can stand for the momentum labels $+$ and $-$, and $\delta_{ij}$ is the usual Kronecker delta \cite{Jordan}.

If the Lorentz transformation $\Lambda$ of Eq.(\ref{transformation}) were a pure rotation, then the associated Wigner rotation would not depend on the momentum of the particles and would therefore have no effect on the entanglement of the states. Furthermore, since every Lorentz transformation can be decomposed into a pure rotation and a pure boost, and pure rotations have no effect on entanglement, we consider in what follows the transformations $\Lambda$ to be pure boosts. On the other hand, we see from Eq.(\ref{Wigner rotation}) that if the Lorentz boost $\Lambda$ is along the direction of propagation of the particles, which we take to be the $z$-axis, then the Wigner rotation becomes the identity transformation, and causes no change in entanglement. Therefore, without loss of generality, we can choose the boost $\Lambda$ to be along the $x$-axis, perpendicular to the direction of propagation of the particles. In this seting, the spin-one representation of the Wigner rotation takes the form
\begin{equation}
\label{spin 1 rotation}
D\left(W\left(\Lambda,p\right)\right)= 	\begin{pmatrix}
	\frac{1}{2}(1+\cos\Omega) & \frac{1}{\sqrt{2}}\sin\Omega & \frac{1}{2}(1-\cos\Omega) \\
	\frac{1}{\sqrt{2}}\sin\Omega & \cos\Omega & \frac{1}{\sqrt{2}}\sin\Omega\\
	\frac{1}{2}(1-\cos\Omega) & \frac{1}{\sqrt{2}}\sin\Omega & \frac{1}{2}(1+\cos\Omega)
	\end{pmatrix},	
\end{equation}  
where $\Omega$ is given by Eq. (\ref{Wigner angle}).

Let $\vert \psi^\prime \rangle = U(\Lambda)\vert \psi \rangle$ denote the state obtained from the original state $\vert \psi \rangle$ after the Lorentz transformation $U(\Lambda)$, i.e., $\vert \psi^\prime \rangle$ is the state described by an observer that is transformed with $\Lambda^{-1}$ with respect to the observer that describes the state with $\vert \psi \rangle$. The initial entanglement calculated from $ \rho = \vert \psi \rangle \langle \psi \vert $ is to be compared with that obtained from $ \rho^\prime = \vert \psi^\prime \rangle \langle \psi^\prime \vert $. As an entanglement measure, we use the linear entropy, defined by
\begin{equation}
\label{linear entropy}
E=\sum_{i} \, 1-Tr\left(\rho_{i}^2\right),	
\end{equation}
where the reduced density matrix $\rho_i$ is obtained from the original density matrix $\rho$ by tracing over all subsystems except the $i$-th. Note that the way a state transforms is independent of the entanglement measure chosen, and we shall see that weakly- or not-entangled states can transform into maximally entangled ones, so that if $E$ is not Lorentz-invariant nor will any other appropriate entanglement measure be. As $E$ is a simpler measure to calculate, we choose to work with this measure (for a whole family of entanglement measures the reader may see \cite{Bengtsson}).

For the initial state (\ref{initial state}) we choose a family of states $\vert p \rangle$ and $\vert s \rangle$ parametrized as follows (this reduces to the parametrization found in \cite{Friis} for the case of spin-$1/2$ particles):
\begin{equation}
\label{momentum state}
\vert p\rangle = \cos\alpha\,\vert p_{+},\, p_{-}\rangle + \sin\alpha \,\vert p_{-},\, p_{+}\rangle,	
\end{equation} 

\begin{subequations}
\label{spin state}
\begin{align}
\vert s_1\rangle=&\sin\theta\cos\phi\vert 1, 1\rangle+\sin\theta\sin\phi\vert 0,0\rangle+\cos\theta\vert -1,-1\rangle \label{parametrization 1}\\
\vert s_2\rangle=&\sin\theta\cos\phi\vert 1, 1\rangle+\sin\theta\sin\phi\vert 0,0\rangle+\cos\theta\vert -1,-1\rangle \label{parametrization 2}\\
\vert s_3
\rangle=&\sin\chi\sin\theta\cos\phi\vert 1, 0\rangle+\sin\chi\sin\theta\sin\phi\vert 0, 1\rangle+\sin\chi\cos\theta\vert 0,-1\rangle+\cos\chi\vert -1,0\rangle \label{parametrization 3}.
\end{align}	
\end{subequations}
The three spin parametrizations cover the eigenstates of the operator $J^2$ in the composite Hilbert space for definite values of the parameters $\theta$, $\phi$, and $\chi$.


\section{Results and discussion}
\label{four}
\noindent In this section we analize the entanglement change for each of the partitions of the complete Hilbert space.

\subsection{Partition A vs. B}

\noindent In agreement with \cite{Alsing}, entanglement is conserved for all states with respect to the $A$ vs. $B$ partition, due to the unitarity of the transformation induced in each of the subspaces $H^{(A)}$ and $H^{(B)}$. That is, since we have $U(\Lambda) = U^{(A)}(\Lambda)\otimes U^{(B)}(\Lambda)$, where $U^{(A)}(\Lambda)$ acts on the $A$ subsystem and $U^{(B)}(\Lambda)$ on the $B$ subsystem, then $\rho \longrightarrow U^{(A)}\left(\Lambda\right)\rho^{(A)} \, U^{\dagger(A)}\left(\Lambda\right)\otimes U^{(B)}\left(\Lambda\right)\rho^{(B)} \, U^{\dagger(B)}\left(\Lambda\right)$ and the linear entropy defined in Eq. (\ref{linear entropy}) remains unchanged when we trace out the $A$ or $B$ degrees of freedom. 

Physically, the unitary character of the $U(\Lambda)$ transformation and the resulting conservation of entanglement is very important, since it is a necesary condition for the results of local measurements to depend, for \textit{every} reference frame, only on the information given by the partial states $\rho^{(A)}$ and $\rho^{(B)}$.  If entanglement failed to be conserved in this case, there would exist different quantum correlations for different inertial observers, in contradiction with the principle of relativity. In this way, every physical process that involves quantum correlations, such as quantum telportation, is independent of the reference frame in which it is analyzed and the experimental results will be the same for every observer. 


\subsection{Partition $p$ vs. $s$}

\noindent Entanglement with respect to the $p$ vs. $s$ partiton \textit{is not} conserved in the general case. This is because of the dependence of the Wigner rotation on the momentum of the particle. If we have an initial momentum superposition, then we have different transformations for the spin part of the system, one for each value of the momentum. Thus, the spin part of the system transforms according to a superposition of Wigner rotations, and the resulting transformed state is no longer a tensor product of spin and momentum states. On the other hand, if the initial momentum state is not a superposition, which means $\cos\alpha = 0$ or $\sin\alpha =0$ in Eq. (\ref{momentum state}), there is only one Wigner rotation for the state; therefore, in this case, the resulting transformed state differs from the original one only by a unitary transformation, which laves the entanglement of the system unchanged.

\begin{figure}[h]
\centering
\scalebox{0.660}{\includegraphics{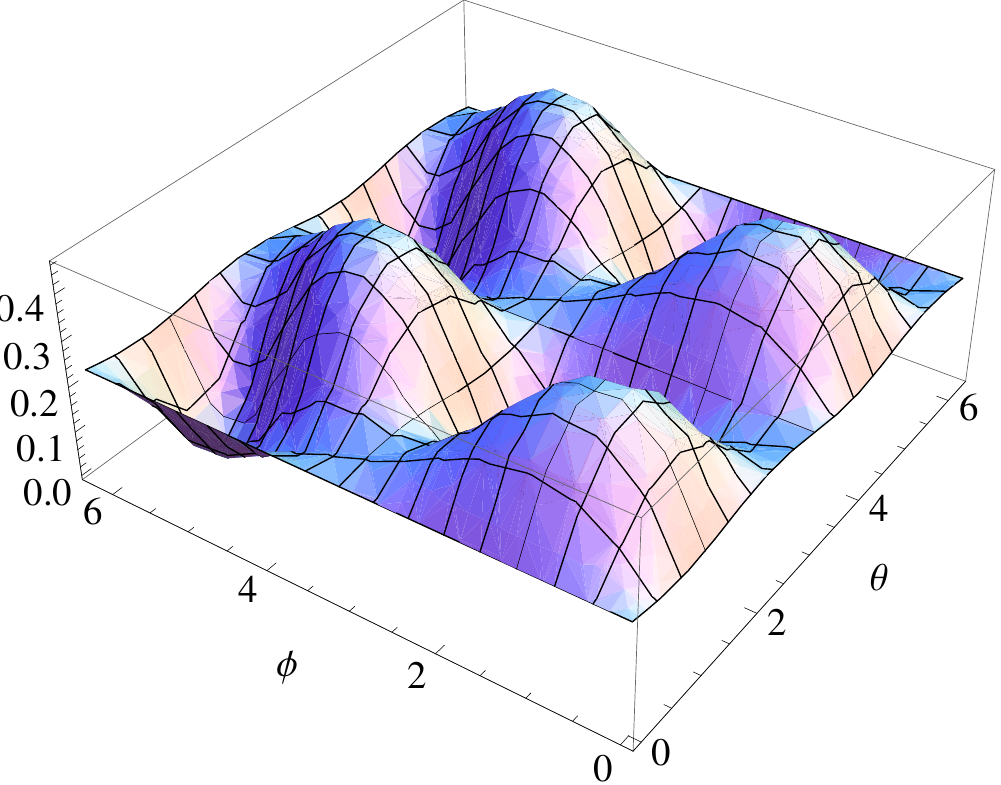}} \quad
\scalebox{0.630}{\includegraphics{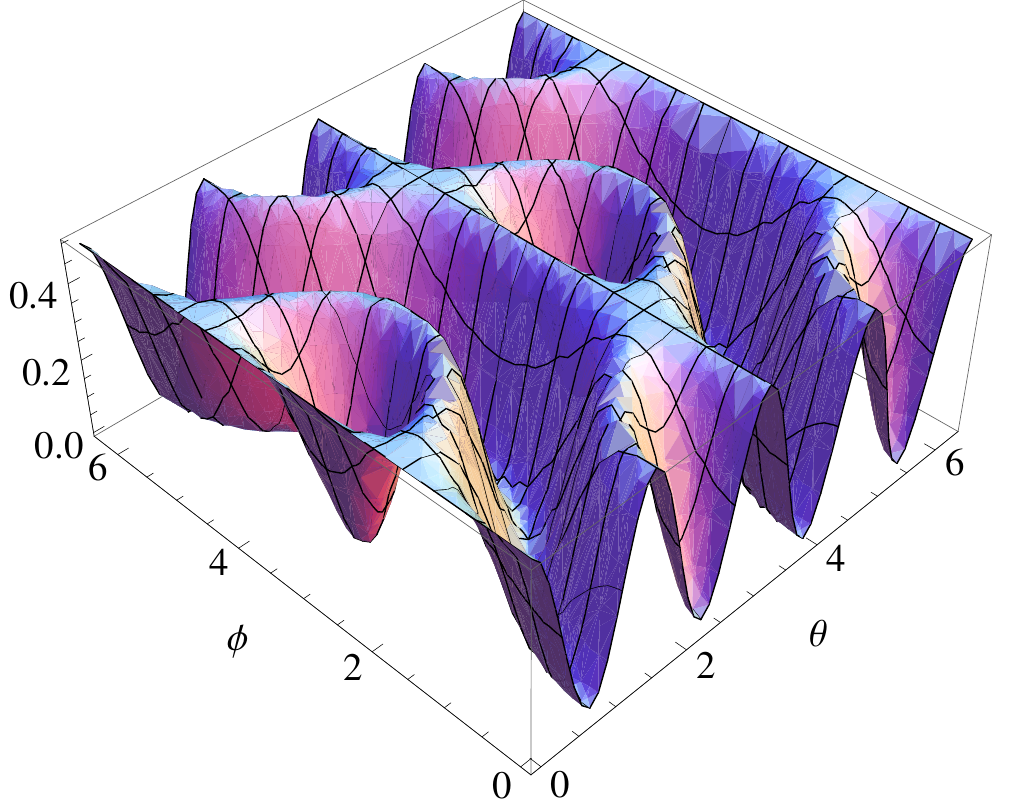}}
\caption{$\Delta E (\theta, \phi)$ for the spin state of Eq. (\ref{parametrization 1}) and partition $p$. vs. $s$. We take $\alpha = \frac{\pi}{4}$. \textbf{Left}: $\Omega = \frac{\pi}{8}$. \textbf{Right}: $\Omega = \frac{\pi}{2}$, corresponding to the limit of the speed of light. The function on the right coincides with that of the spin-$\frac{1}{2}$ case.}
\label{par1partspmax}
\end{figure}

The change in linear entropy depends strongly on the Wigner angle, as can be seen from Fig. \ref{par1partspmax}. In this figure we plot the entanglement change $\Delta E$ as a function of $theta$ and $phi$ for different values of $\Omega$. We have chosen values of $\Omega$ that differ considerably to stress the effect that this parameter has on the shape of the surface. When both the speed of the particles and the speed of the boost $\Lambda$ approach the speed of light we have $\Omega = \pi/2$.

Interestingly, in the limit of the speed of light ($\Omega = \pi/2$), the entanglement change of the spin state given by Eq. (\ref{parametrization 1}) produces exactly the same function of $\theta$ and $\phi$ as that obtained by Friis et al in \cite{Friis} for the spin-$1/2$ composite state $\sin\theta\cos\phi\vert 0,0\rangle +\sin\theta\sin\phi\left(\frac{\vert0,0\rangle+\vert 1,1\rangle}{\sqrt{2}}\right) +\cos\theta\vert 1,1\rangle$. In both parts of Fig. \ref{par1partspmax} there are minima that correspond to the maximally entangled state $\frac{1}{\sqrt{3}}(\vert 1,1 \rangle-\vert 0,0 \rangle +\vert -1,-1 \rangle)$.

In all cases, the dependence of $\Delta E $ on the parameter $\alpha$ is merely a change of scale, which is zero for non-superposed momentum states and is maximal for the initial momentum state $\frac{1}{\sqrt{2}}(\vert p_+, p_- \rangle +\vert p_-, p_+ \rangle)$. For this reason, in what follows we consider only the case $\alpha = \pi/4$ without loss of generality.

It is important to point out that the entanglement change is due to the linear superposition of initial momentum states and not due to the entnglement of the initial state $\vert p\rangle$. To see this, let us start with the separable state $\vert p \rangle = \frac{1}{\sqrt{2}}\vert p_+\rangle\left(\vert p_+ \rangle + \vert p_-\rangle \right)$. This state is not of the form of Eq. (\ref{momentum state}) but we use it here for the sake of argument. With this choice of initial momentum superposition, any state of the form $\vert p \rangle \otimes \vert s \rangle$, with $\vert s \rangle$ being an arbitrary initial spin state, transforms into $  \frac{1}{\sqrt{2}}\vert\Lambda p_+,\Lambda p_+\rangle U_s\left(p_+,p_+\right)\vert s \rangle + \frac{1}{\sqrt{2}}\vert \Lambda p_+,\Lambda p_-\rangle U_s\left(p_+,p_-\right)\vert s \rangle$, where $U_s(p,q)$ is the transformation induced by the boost in the two-particle spin subspace when the first particle has momentum $p$ and the second has momentum $q$. This final state is entangled in the spin and momentum degrees of freedom since, in general, the transformations on spin space $U_s\left(p_+,p_+\right)$ and  $U_s\left(p_+,p_-\right)$ are not equal.

\begin{figure}[h]
\scalebox{0.750}{\includegraphics{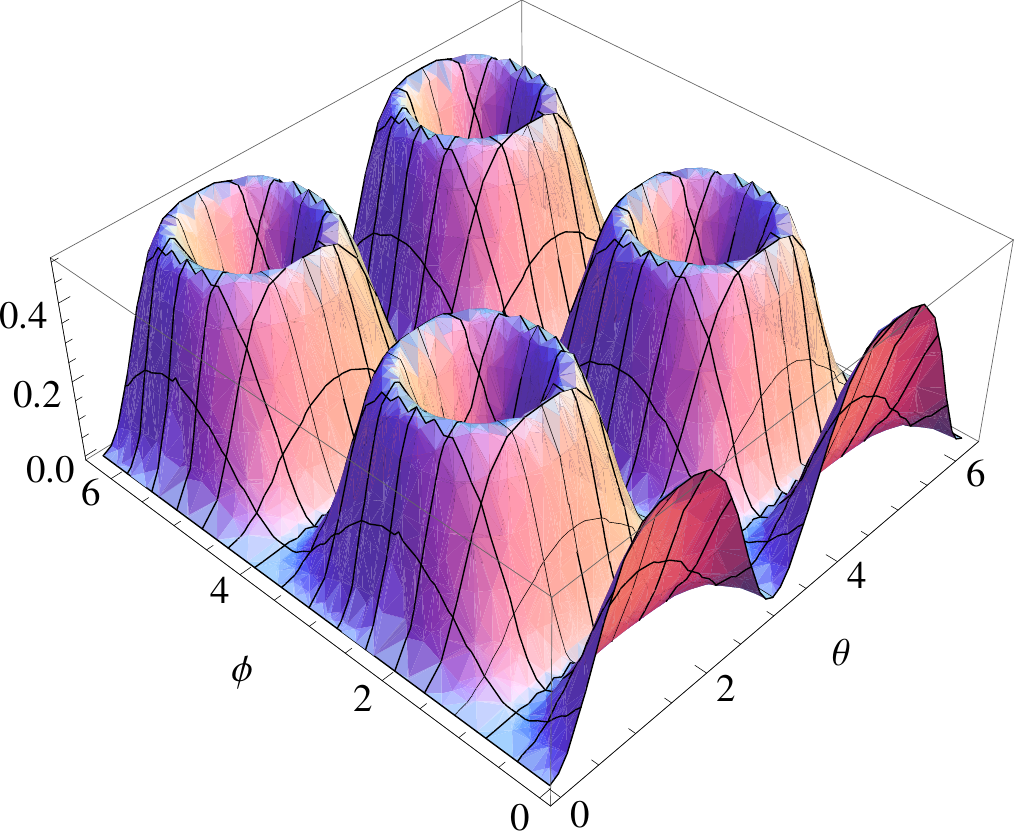}}
\scalebox{0.750}{\includegraphics{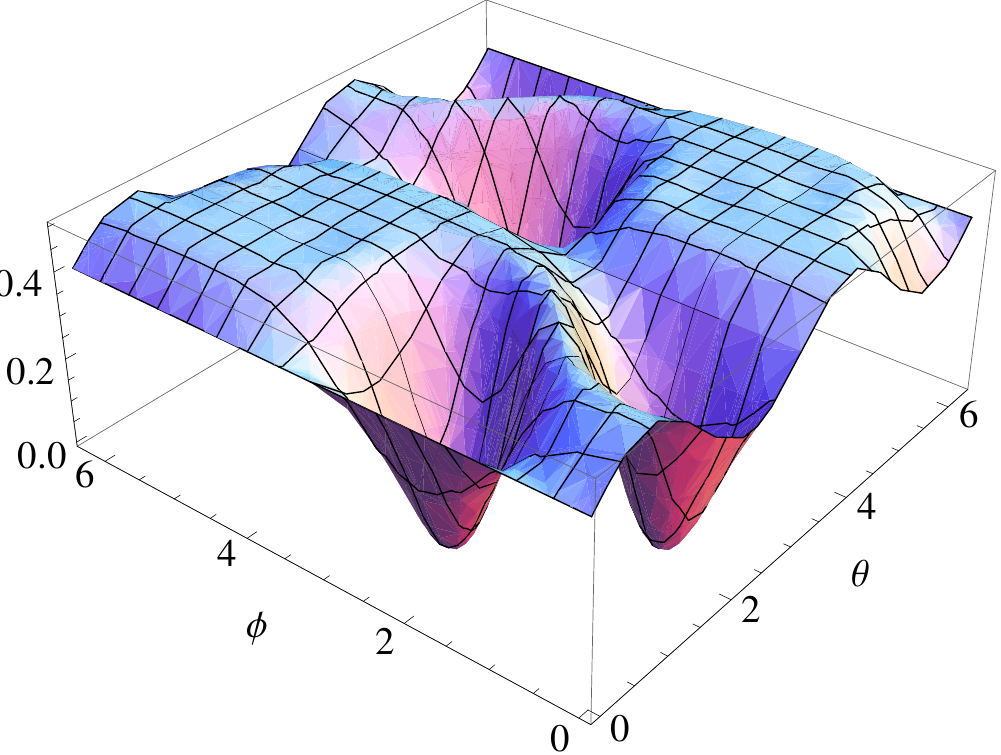}}
\caption{Left: $\Delta E(\theta,\phi)$ for the spin state of Eq. (\ref{parametrization 2}) and partition $p$ vs. $s$, with $\Omega = \frac{\pi}{2}$. The points in the center of the depressions correspond to the Bell-type state $\frac{1}{\sqrt{2}}(\vert 1,-1 \rangle-\vert -1,1 \rangle)$. Right: Same, for parametrization (\ref{parametrization 3}) and $\chi=2\pi/3$; in this case the minima correspond to the state $\vert s\rangle = \frac{1}{2}\,\left( \vert 1,0\rangle + \vert 0,1\rangle - \vert 0,-1\rangle - \vert -1,0\rangle \right)$.}
\label{par2partspograndemax}
\end{figure}

We now return to our discussion of the $p$ vs. $s$ partition. Fig. \ref{par2partspograndemax} (left) shows the entanglement change for the spin initial state given by Eq. (\ref{parametrization 2}) and a Wigner angle $\Omega = \pi/4$. The minima of the surface correspond to the Bell-type state ${\frac{1}{\sqrt{2}}(\vert 1,-1 \rangle-\vert -1,1 \rangle)}$, which undergoes no entropy change for this particular Wigner angle. When the parametrization given by Eq. (\ref{parametrization 3}) is used (Figure \ref{par2partspograndemax} (right)) it is possible to generate entanglement for a large number of states under a Lorentz transformation with the same Wigner angle of $\pi/4$. Here we have used $\chi=2\pi/3$. The invariant state (minima of plot) is $\vert s\rangle = \frac{1}{2}\,\left( \vert 1,0\rangle + \vert 0,1\rangle - \vert 0,-1\rangle - \vert -1,0\rangle \right)$.

\subsection{Partition $1$ vs. $3$}

\begin{figure}[h]
\scalebox{0.63}{\includegraphics{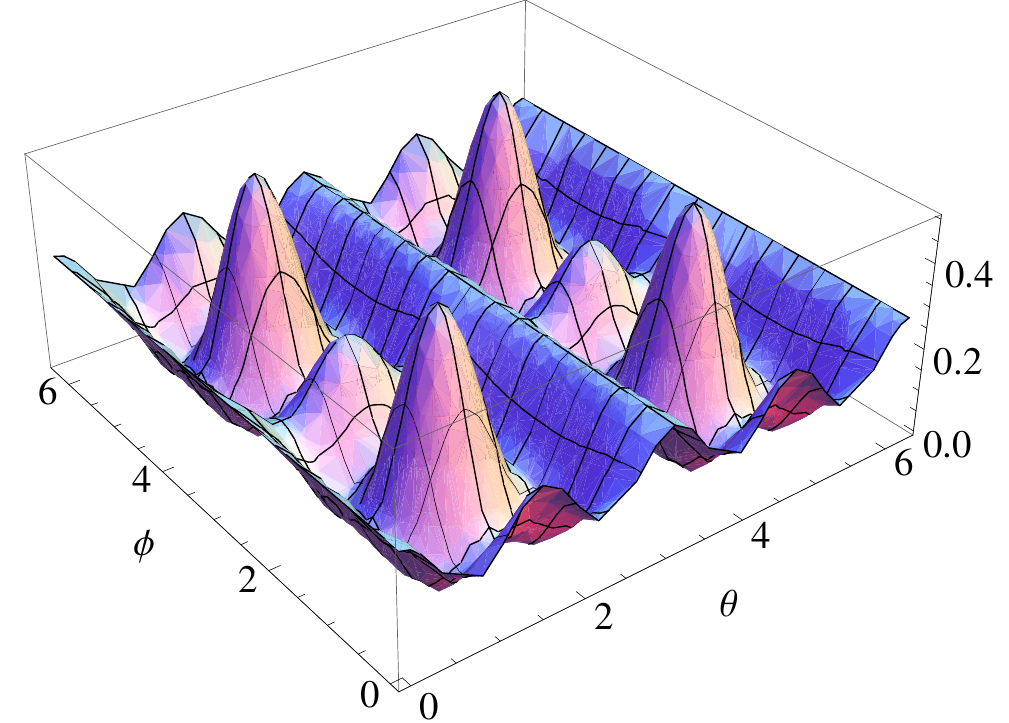}} \quad
\scalebox{0.64}{\includegraphics{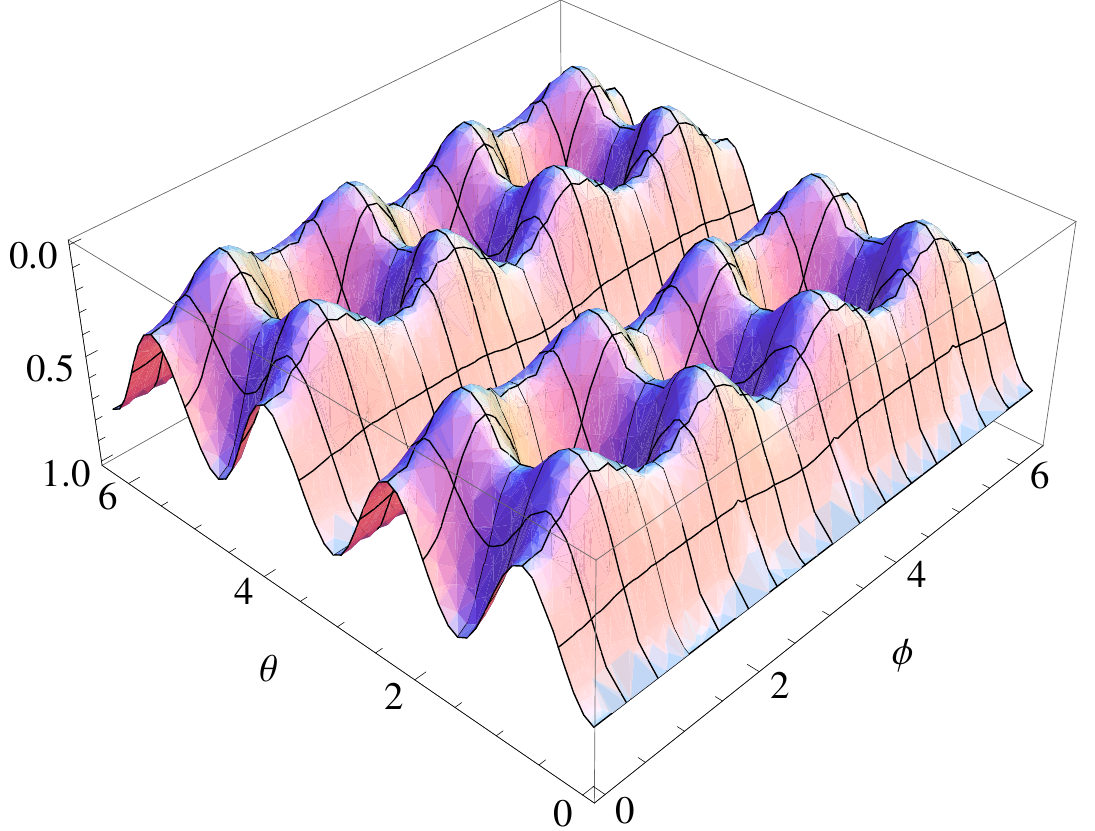}} \\
\centering
\scalebox{0.63}{\includegraphics{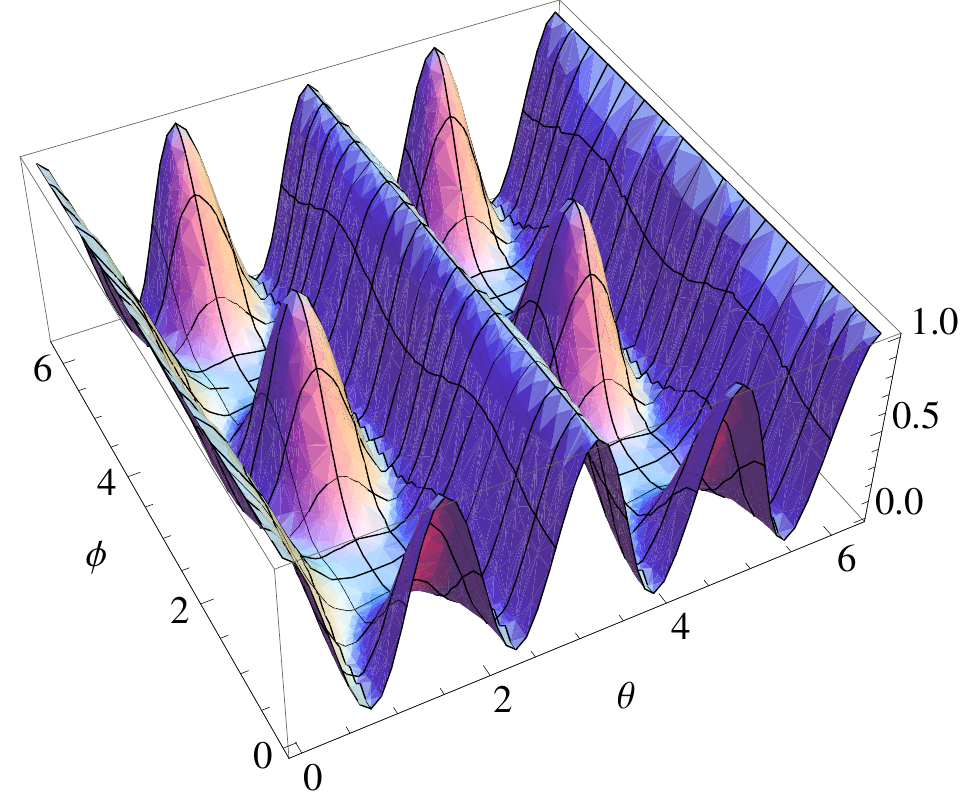}}
\caption{ $\Delta E(\theta,\phi)$ for the spin state of Eq. (\ref{parametrization 1}) and partition $1$ vs. $3$. \textbf{Above:} $\Omega = \frac{\pi}{8}$. In this case the maxima correspond to the state $\vert 0, 0 \rangle$ (\textbf{left}), and the minima to the states $\frac{1}{\sqrt{3}}(\vert 1, 1 \rangle \pm \vert 0, 0 \rangle)+\vert -1, -1 \rangle$. \textbf{Below:} $\Omega = \frac{\pi}{2}$. Maxima belong to the states $\vert 1, 1 \rangle$ and $\vert -1, -1 \rangle$.}
\label{par1part1}
\end{figure}

\noindent As in the previous case, entanglement is \textit{not} conserved for all states in this type of decomposition and the maxima and minima of $\Delta E$ vary significantly with the Wigner angle $\Omega$. To illustrate this point, we take the state given in Eq. (\ref{parametrization 1}). In Fig. \ref{par1part1} we show the entanglement change, as a function of $\theta$ and $\phi$, for the values values of  $\Omega = \pi/8$ and $\Omega = \pi/2$. When the Wigner angle equals $\pi/8$, i.e.``small" velocities, the entanglement change is maximum for the state $\vert 0 0 \rangle$, while the states $\vert 1, 1 \rangle$ and $\vert -1, -1 \rangle$ correspond to local maxima. Above on the right of Fig. \ref{par1part1} is shown the same function as that on the left side but turned upside down, in order to look at the minima. Such minima correspond to the maximally entangled states $\frac{1}{\sqrt{3}}(\vert 1,1\rangle \pm \vert 0,0\rangle + \vert -1,-1\rangle)$. In spite that for this partition both of these states conserve entanglement for all values of the Wigner angle, only the state $\frac{1}{\sqrt{3}}(\vert 1,1\rangle - \vert 0,0\rangle + \vert -1,-1\rangle)$ is invariant under transformations of the form $U^{(A)}_s\left(\Omega\right)\otimes U^{(B)}_s\left(-\Omega\right)$, that are, as we will show later, the kind of maps induced on the spin space by the Lorentz boost. In this notation, $U^{(i)}_s(\Omega)$ stands for the transformation given by Eq. (\ref{spin 1 rotation}) and acts on the $H^{(i)}_s$ subspace. This invariance appears more clearly when we consider the $p$ vs. $s$ partition, in which $\frac{1}{\sqrt{3}}(\vert 1,1\rangle - \vert 0,0\rangle + \vert -1,-1\rangle)$ has zero entanglement change for all values of $\Omega$, but $\frac{1}{\sqrt{3}}(\vert 1,1\rangle + \vert 0,0\rangle + \vert -1,-1\rangle)$ \textit{does} generate a change in entanglement for some values of the Wigner angle.

On the other hand, for $\Omega = \pi/2$ (Fig. \ref{par1part1}, below), the states $\vert 1,1\rangle$ and $\vert -1,-1\rangle$ have a maximal change in linear entropy, while the state $\vert 0,0 \rangle$ stays now at the bottom of the plot. We see, thus, how the velocity of the particles, as well as the rapidity of the Lorentz boost, play an important role in the entanglement change of the state.

\begin{figure}[h]
\scalebox{0.70}{\includegraphics{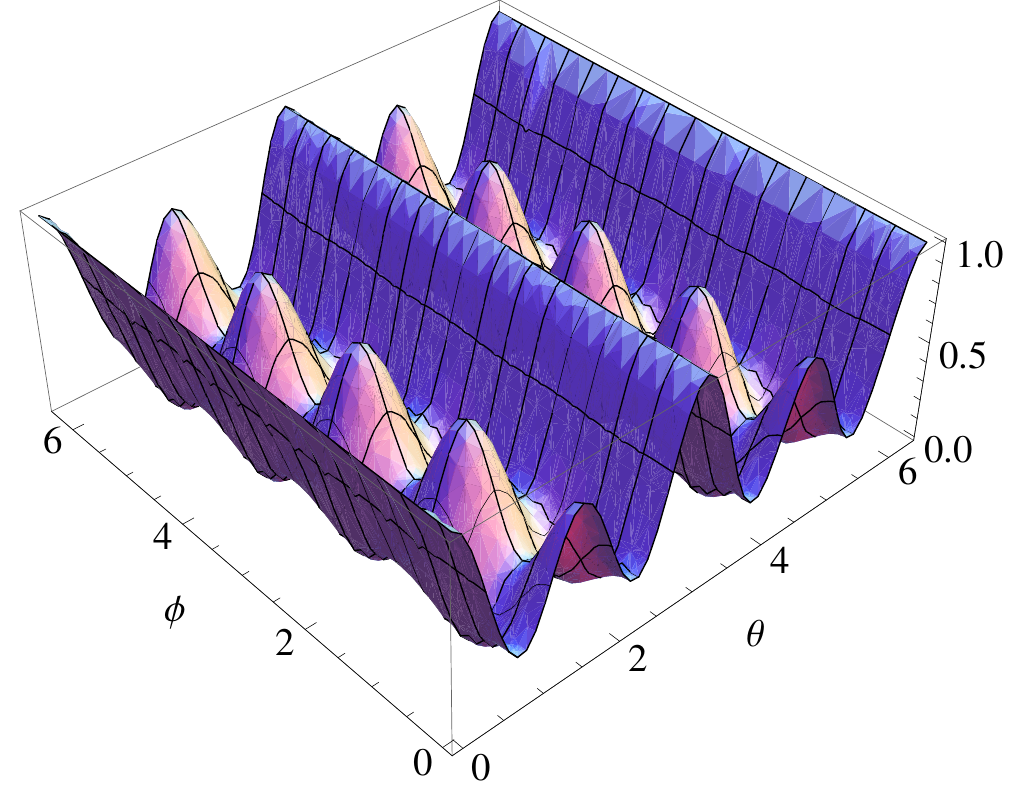}} \quad 
\scalebox{0.66}{\includegraphics{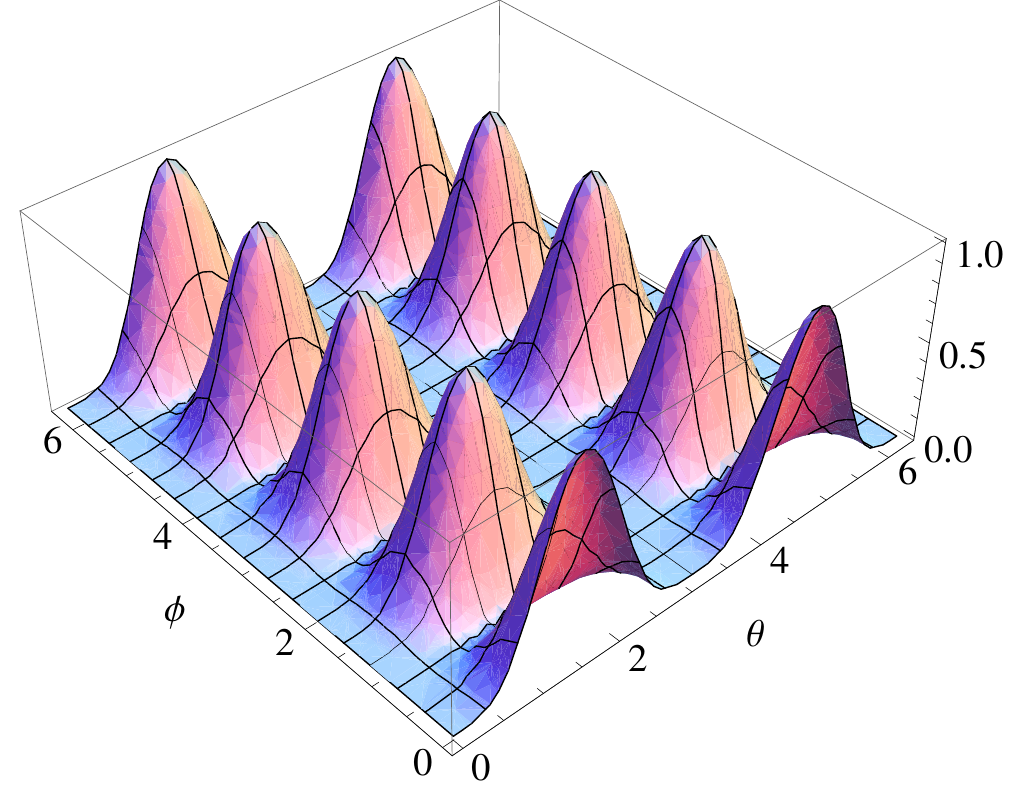}}
\caption{ $\Delta E(\theta,\phi)$ for the spin state of Eq. (\ref{parametrization 2}) and partition $1$ vs. $3$. \textbf{Left:} $\Omega = \frac{\pi}{4}$. Maxima correspond to the state $\vert 0,0 \rangle$. \textbf{Right:}$\Omega = \frac{\pi}{2}$. Maxima belong to the states $\vert 1, -1 \rangle$ and $\vert -1, 1 \rangle$. In both cases entanglement is conserved for the invariant state $\frac{1}{\sqrt{3}}(\vert 1,-1\rangle + \vert 0,0\rangle + \vert -1, 1\rangle)$}
\label{par2part1max}
\end{figure}

For the second parametrization, given in Eq. (\ref{parametrization 2}), the linear entropy behaves in a similar way as in the first parametrization. As can be seen from Fig. \ref{par2part1max}, the states $\vert1,-1\rangle$ and $\vert -1,1 \rangle$, have a smaller entanglement change than the state $\vert 0,0 \rangle$ for a Wigner angle of $\Omega = \pi/4$. We know from the last case that $\vert 0,0 \rangle$ has a maximum entanglemet change for $\Omega = \pi/8$. As the Wigner angle approaches $\pi/2$, the change in linear entropy corresponding to the initial spin state $\vert 0,0 \rangle$ decreases, while that corresponding to $\vert 1,-1\rangle$ and $\vert -1, 1 \rangle$ increases continuously. Finally, in the limit of the speed of light, $\Omega = \pi /2$, the change in linear entropy corresponding to the state $\vert 0,0 \rangle$ vanishes (as we already know from the previous section), and the states $\vert 1,-1\rangle$ and $\vert -1, 1 \rangle$ (both with momentum part given by Eq. (\ref{momentum state}) with $\alpha = \pi/4$) become, after the Lorentz boost, maximally entangled states with respect to the $1$ vs. $3$ partition. It is interesting to note that the state $\frac{1}{\sqrt{3}}(\vert 1-1\rangle + \vert 00 \rangle + \vert -11\rangle)$ is invariant under the maps induced by the Lorentz boost studied here, as is the case of the state $\frac{1}{\sqrt{3}}(\vert 11\rangle + \vert 00 \rangle + \vert -1-1\rangle)$ studied in the last section. Both of these spin states are maximally entangled and, as they remain unchanged under the Lorentz transformations presented here, they are interesting for quantum information purposes in an EPR-like relativistic framework.   


\subsection{Invariant states}

\noindent In the physical situation considered here the boost direction is kept fixed and the particles momenta are always opposite in relation to each other (and given by sharp distributions). These two facts imply the that there exists a complete set of states in the two-particle spin Hilbert space such that, for each state in this set, the action of the Lorentz boost is merely a multiplication by a global phase factor. To see this, first we note that the Wigner rotation for a single particle is restricted to be an element of $SO(2)$, the rotation group in the plane, since a fixed boost direction implies a fixed axis of rotation. Moreover, since the momenta of the particles are always opposite, by Eq. (\ref{Wigner angle}) the angle of rotation for one particle will be of the same magnitude but opposite sign as that of the other particle. Then, the transformation on the two-particle spin space induced by the Lorentz transformation will be of the form $U_s\left(\Omega\right) = U^{(A)}_s\left(\Omega\right)\otimes U^{(B)}_s\left(-\Omega\right)$. 

Now we show that the set
\begin{equation}
\{U_s\left(\Omega\right) = U^{(A)}_s\left(\Omega\right)\otimes U^{(B)}_s\left(-\Omega\right) \ \vert \ 0\leq \Omega < 2\pi\}
\end{equation}
forms a representation of $SO(2)$. The product of two elements $(\Omega_1)$ and $(\Omega_2)$ $\in$ $SO(2)$ is given by the sum of the parameters that define these elements
\begin{equation}
\left(\Omega_1\right)\left(\Omega_2\right) = \left(\Omega_1+\Omega_2\right).
\end{equation}
Thus,
\begin{align*}
U_s\left(\left(\Omega_1\right)\left(\Omega_2\right)\right) = & \, U_s\left(\Omega_1+\Omega_2\right)\\
= & \, U^{(A)}_s\left(\Omega_1+\Omega_2\right)\otimes U^{(B)}_s\left(-\Omega_1-\Omega_2\right) \\
= & \, U^{(A)}_s\left(\Omega_1\right) U^{(A)}_s\left(\Omega_2\right)\otimes U^{(B)}_s\left(-\Omega_1\right)  U^{(B)}_s\left(-\Omega_2\right) \\
= & \, \left(U^{(A)}_s\left(\Omega_1\right)\otimes U^{(B)}_s\left(-\Omega_1\right) \right)\left(U^{(A)}_s\left(\Omega_2\right)\otimes U^{(B)}_s\left(-\Omega_2\right)\right)\\
= & \, U_s\left(\Omega_1\right)U_s\left(\Omega_2\right),
\end{align*}
which proves our claim. Since the representations of $SO(2)$ are completely reducible we can, by means of a change of basis, reduce the transformation $U_s(\Omega)$, for any $\Omega$, in terms of the irreducible representations of $SO(2)$, which are of the form $e^{im\Omega}$, with $\Omega \in \mathbb{Z}$. For spin-one systems, the transformation $U_s(\Omega)$ takes the form
\begin{equation}
U_{s}\left(\Omega\right) \longrightarrow\begin{pmatrix}
e^{2i\Omega} & 0 & 0 & 0 & 0 & 0 & 0 & 0 & 0  \\
0 & e^{-2i\Omega}& 0 & 0 & 0 & 0 & 0 & 0 & 0 \\
0 & 0 & e^{i\Omega} & 0 & 0 & 0 & 0 & 0 & 0\\
0 & 0 & 0 & e^{-i\Omega} & 0 & 0 & 0 & 0 & 0\\
0 & 0 & 0 & 0 & e^{i\Omega} & 0 & 0 & 0 & 0\\
0 & 0 & 0 & 0 & 0 & e^{-i\Omega} & 0 & 0 & 0\\ 
0 & 0 & 0 & 0 & 0 & 0 & 1 & 0 & 0\\
0 & 0 & 0 & 0 & 0 & 0 & 0 & 1 & 0\\
0 & 0 & 0 & 0 & 0 & 0 & 0 & 0 & 1\\
\end{pmatrix}.
\end{equation}
Therefore, each element of the basis which diagonalizes $U_s$ changes under a Lorentz boost only by a global phase factor. In this sense, such states may be called invariant under the kind of Lorentz boosts considered in this work (always perpendicular to the momenta of the particles). 
Moreover, entanglement is conserved for every linear combination of states that transform under the same irreducible representation of $SO(2)$, and all the information about the entanglement change of an arbitrary state is kept in the relative phase factors $e^{im\Omega}$.

For particles with arbitrary spin, the transformation $U_s$ is also diagonal. The multiplicity $a_m$ of a given irreducible representation $D^{(m)}$, that is, the number of times this representation appears in the (reducible) representation element $U_s$, can be calculated as \cite{Tung}: 
\begin{equation}
\label{multiplicity}
a_m = \frac{1}{\vert\mathcal{G}\vert}\sum_{g\in \mathcal{G}}\, \chi(g) \left(\chi^{(m)}(g)\right)^*,
\end{equation}
where $\vert\mathcal{G}\vert$ is the order of the group $\mathcal{G}$, $\chi(g)$ is the character of the element 
$g \in \mathcal{G}$ for an arbitrary representation, and $\chi^{(m)}(g)$ is the character of the element 
$g$ for the irreducible representation labeled by $m$. The sum is taken over all group elements and becomes an integral for the case of continous groups. For the group $SO(2)$, $\frac{1}{\vert\mathcal{G}\vert}\sum_{g\in \mathcal{G}}=\frac{1}{2\pi}\int_0^{2\pi}d\Omega$. For a two spin-$j$ paticle system 
\begin{equation}
\label{tensor character}
\chi(\Omega) = \chi^{(A)}(\Omega)\chi^{(B)}(\Omega),	
\end{equation}   
where the characters for the single-particle space transformations, $\chi^{(A)}(\Omega)$ and $\chi^{(B)}(\Omega)$, are both given by \cite{Tung}:
\begin{equation}
\label{character}
\chi^{(A)}{\Omega} = \chi^{(B)}{\Omega} = \sum^j_{m=-j}\, e^{-im\Omega}.
\end{equation}
Putting together Eqs. (\ref{multiplicity}), (\ref{tensor character}) and (\ref{character}) yields, after a short calculation,
\begin{equation}
a_m = 2j+1-\vert m\vert,
\end{equation}
so that for an arbitrary spin $j$ we have
\begin{equation}
U_s(\Omega) \longrightarrow diag (\underbrace{e^{2ij\Omega}}_{a_{2j} \ times},\underbrace{e^{i(2j-1)\Omega}}_{a_{2j-1} \ times},\cdot\cdot\cdot, \underbrace{e^{-2ij\Omega}}_{a_{-2j} \ times}).
\end{equation}

Finally, and based on this last result, we note that the invariance of entanglement dos not depend, in general,
on the initial entanglement of the spin state, but on its transformation properties. For example, the state 
\begin{align}
\vert s \rangle =& \, \frac{1}{2}\left(\vert 1, 1 \rangle + \vert 1, -1 \rangle + \vert -1, 1 \rangle + \vert -1,-1 \rangle \right) \nonumber \\
= & \, \frac{1}{2}\left(\vert 1 \rangle +\vert -1 \rangle \right)\otimes\left(\vert 1\rangle +\vert -1 \rangle \right)
\end{align}
is separable and also invariant under the $U_s$ transformations.


\section{conclusions}
\label{five}
\noindent The entanglement with respect to the $A$ vs. $B$ partition is invariant under a Lorentz transformation. This fact is a direct consequence of the unitarity of the transformation $U(\Lambda) = U^{(A)}(\Lambda)\otimes U^{(B)}(\Lambda)$. The conservation of this kind of entanglement is fundamental for the consistency between quantum-mechanical predictions and relativistic transformations.

On the other hand, entanglement is not conserved neither for the $p$ vs. $s$ partition nor for the $1$ vs. $3$ partition. This is due to the momentum-dependence of the Wigner rotation, which induces, for momentum-superposed initial states, different transformations for different particles. The entanglement change, for a given $\Omega$, reaches a maximum for a homogeneous momentum superposition ($\alpha = \pi/4$), while it vanishes for non-superposed momentum states.

The dependence of the linear entropy change with respect to the initial spin state and to the Wigner angle $\Omega$ is more interesting. Generally, the states for which the entanglement change is greater are separable spin states, while this quantity remains constant (normally) for states with maximal entanglement. However, the initial spin entanglement is not a crucial factor regarding the entanglement change after a Lorentz boost, since we can find separable spin states which conserve entanglement with respect to all partitions and all values of the Wigner angle. In the same way, there exist maximally entangled spin states
that do not conserve entanglement with respect to the $p$ vs. $s$ partition.

The transformation properties of the spin states is the crucial fact that determines the change in entanglement. More precisely, what matters is the way in which the spin states transform under operations of the form $U_s\left(\Omega\right) = U^{(A)}_s\left(\Omega\right)\otimes U^{(B)}_s\left(-\Omega\right)$, that are representations of the group $SO(2)$. In particular, the spin states that are invariant under this type of transformations conserve entanglement for every partition and all values of the Wigner angle, with no regard to the initial momentum state. Since the action of the group $SO(2)$ is reducible in the spin state space, we can find an orthonormal basis of this space such that each state in the basis transforms, under a Lorentz boost, only by multiplication of a factor of the form $e^{im\Omega}$. In this basis, the Wigner rotation takes a simple form and all the information about the entanglement change lies in the relative phases $e^{i(m-m^\prime)\Omega}$, for $m$, $m^\prime$ $\in \mathbb{Z}$. All linear combinations of states that transform under the same irreducible representation of $SO(2)$ are invariant under the kind of Lorentz boosts considered in this work and therefore conserve their entanglement. This result can be generalized to systems of arbitrary spin $j$ so that, in principle, invariant subspaces of an arbitrary dimension can be constructed, considering the adequate type of particles. This fact may be useful for quantum information processes in which relativistic considerations are relevant.




\section{Acknowledgments}
This work was partially supported by DGAPA-UNAM under project IN102811.

\end{document}